\documentclass[aps,prl,twocolumn,preprintnumbers,amsmath,amssymb]{revtex4}

\usepackage{graphicx}
\usepackage{dcolumn}
\usepackage{bm}
\usepackage{float}

\newcommand{\tOuter}{\ensuremath{\tilde{t}}}
\newcommand{\tInner}{\ensuremath{t}}

\begin{document}

\title{Nanoscale ferromagnetism in non-magnetic doped semiconductors}

\author{Erik Nielsen$^1$}
\author{R. N. Bhatt$^{1,2}$}
\affiliation{$^1$Department of Electrical Engineering, Princeton University, Princeton, NJ 08544-5263}
\affiliation{$^2$Princeton Center for Theoretical Physics, Jadwin Hall, Princeton, NJ 08544}

\date{\today}

\begin{abstract}
While ferromagnetism at relatively high temperatures is seen in diluted magnetic semiconductors such as Ga$_{1-x}$Mn$_x$As, doped semiconductors without magnetic ions have not shown evidence for ferromagnetism.  Using a generalized disordered Hubbard model designed to characterize hydrogenic centers in semiconductors, we find that such systems may also exhibit a ferromagnetic ground state, at least on the nanoscale. This is found most clearly in a regime inaccessible to bulk systems, but attainable in quantum dots as well as heterostructures. We present numerical results demonstrating the occurrence of high spin ground states in both lattice and positionally disordered systems.  We examine how the magnetic phases are affected by characteristics of real doped semiconductors, such as positional disorder and electron-hole asymmetry. 
\end{abstract}

\pacs{71.23.-k, 71.27.+a, 73.21.La, 75.40.Mg, 75.50.Pp, 75.75.+a}

\maketitle

The Hubbard model\cite{HubbardOriginal} is perhaps the simplest model containing the essential character of correlated electrons; it consists of tight binding on a lattice with a purely on-site electron-electron repulsion.  It has been studied extensively, \emph{e.g.}, in the large-$U$ regime\cite{largeU}, on different lattices\cite{diffLatt}, with multiple possibly degenerate bands\cite{multiBand}, and with binary alloy disorder\cite{Byczuk_2003}.  It has been used to model Mott-insulator oxides \cite{MottBook}, high-$\mathrm{T}_c$ superconductors \cite{Cuprates}, organic conductors \cite{organicPapers} as well as hydrogenic centers in doped semiconductors\cite{dopedSemi}.  For the latter, it is especially relevant in the insulating phase, where the Coulomb interaction is large compared to the kinetic energy.  

The Hubbard model on a lattice is defined by the Hamiltonian:
\begin{equation}
\mathcal{H} = -t\sum_{\langle i,j\rangle\sigma} \left( c^\dag_{i\sigma} c_{j\sigma} + \mbox{c.c} \right) + U\sum_i n_{i\uparrow}n_{i\downarrow} \label{eqnHubHamOriginal}
\end{equation}
where $t$ and $U$ are the kinetic and Coulomb energy parameters respectively, $c^\dag_{i\sigma}$ ($ c_{i\sigma}$) is the usual electron creation (annihilation) operator on site $i$ with spin $\sigma$, and angular brackets denote nearest neighbors.  For example, in the tight-binding model with hydrogenic wavefunctions, $t(r)=2(1+r/a_{\mathrm{B}})\exp(-r/a_{\mathrm{B}})$\cite{Bhatt_1981}.  Nagaoka\cite{Nagaoka} showed that in the limit $U/t\rightarrow\infty$, the Hubbard model on a finite bipartite lattice of dimension $d\ge 2$ with periodic boundary conditions and a single hole (away from half filling), has a ferromagnetic ground state.  The reason for this can be understood by considering a system with hole density $\delta$. The kinetic energy gain of the holes is more restricted in an antiferromagnetic background of spins than in a fully spin-polarized background\cite{BrinkmanRiceShraimonSiggia}.  This restriction is of order $\delta t$, whereas the antiferromagnetic superexchange cost is order $J=4t^2/U$\cite{AFhalfFilling}.  At small $\delta$ and large enough $U$, $\delta t > J$, and the system prefers a ferromagnetic configuration because it allows for less confined carriers.  Subsequent work to Nagaoka's pioneering result has shown that ferromagnetism is a subtle effect depending on lattice geometry.  For example, Lieb and Mattis\cite{LiebMattis} proved that in finite one-dimensional systems with zero-wavefunction or zero-derivative boundary conditions, the ground state must be a singlet, and Haerter and Shastry\cite{ShastryAFTriangle} recently showed that on the frustrated triangular lattice ($t<0$), an itinerant hole actually helps to produce an \emph{antiferromagnetic} ground state. 

\begin{figure}[H]
\begin{center}
\includegraphics[width=2.8in]{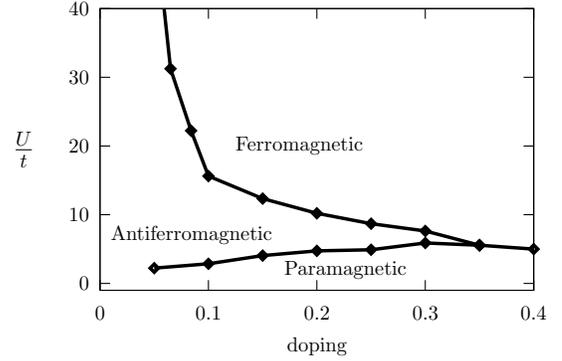}
\caption{Zero temperature mean-field theory phase diagram of the Hubbard model on a simple cubic lattice (512 sites).
\label{figMFTdiagram}}
\end{center}
\end{figure} 

The mean field diagram of this model on a simple cubic lattice is given in figure \ref{figMFTdiagram}, which agrees qualitatively with the more extensive work in two dimensions by Hirsch\cite{HirschMFT}.  An antiferromagnetic phase exists at half-filling for all values of $U/t$, due to the the effective antiferromagnetic interaction from superexchange.  Such a mean-field analysis does not include the possibility of phase separation, {\it e.g.}~the existence of polarons corresponding to ``carrier-rich" ferromagnetic and ``carrier-poor" antiferromagnetic spatial regions.  Phase separation may alter the simple phase diagram of fig.~\ref{figMFTdiagram} substantially\cite{EisenbergHuseAltshuler}, though Dagotto et~al.\cite{DagottoPhaseSep} argue for its absence in the Hubbard model based on 10- and 16-site square lattices.  Even though its precise location in phase space depends on dimension and requires more careful work, it is clear that for large enough $U/t$ we expect an antiferromagnetic to ferromagnetic transition on some mesoscopic or macroscopic length scale, as a function of doping.

However, in spite this expectation, Nagaoka ferromagnetism does not seem to have been seen experimentally.  Whereas most Mott-insulator oxides appear not to have a large enough $U/t$ ratio to allow for it, doped semiconductors have $U/t$ tunable over several orders of magnitude (due to the exponential dependence of the hopping $t$ on the dopant spacing), and thus are a promising candidate for Nagaoka ferromagnetism.  Unfortunately, the dopants are not arranged on a superlattice, and the Hubbard Hamiltonian must be modified to include a site-dependent hopping term.
\begin{equation}
\mathcal{H} = - \sum_{\langle i,j\rangle\sigma} \left( t_{ij}c^\dag_{i\sigma} c_{j\sigma} + \mbox{c.c} \right) + U\sum_i n_{i\uparrow}n_{i\downarrow} \label{eqnHubHamDisordered}
\end{equation}
It has been shown\cite{BhattLee} that in uncompensated semiconductors (half-filled Hubbard band), the randomness of the dopants results in a valence-bond glass (random singlet) phase being a better description of the ground state than the antiferromagnet predicted on a lattice.  Experiments on compensated systems (away from half-filling), show no evidence of ferromagnetism in conventional semiconductors with non-magnetic dopants\cite{Hirsch_NoExpFerro}.  This can be attributed to the localization of the holes due to the strong randomness in hopping parameters that results from random dopant positions.  The holes are consequently unable to gain the kinetic energy which favors a spin-polarized background.  Thus, despite the ability to tune $U/t$ over such a large range, Nagaoka ferromagnetism remains elusive.

In this work, we show that there exists a regime of doped semiconductor systems that is attainable in nanoscale quantum dots and heterostructures, but not accessible to bulk systems, which is more suited to the occurrence of Nagaoka type ferromagnetism. In this region, ferromagnetism is found at least at the nanoscale, and has a higher likelihood of emerging on macroscopic scales ({\it e.g.}~in modulation doped systems).

The existence of such a regime is suggested by special properties of the hydrogen atom (and hydrogenic centers) that effectively reduce the disorder and move the randomly doped semiconductor in the direction of a lattice problem.  The key property is that the second electron of a $H^-$ ion is bound by only 0.0555 Ryd\cite{BS_QMbook}, an energy much less than the 1 Ryd binding of the initial electron.  This property is linked with the fact that the two-electron wavefunction of $H^-$ is spatially much larger than the 1s wavefunction of the hydrogen atom.  It is much easier for the second electron on a hydrogenic center to hop away than it is for a single electron on such a center to do so. Thus, near half filling, in Hubbard model parlance, the hopping amplitude for an electron is much larger than for a hole.  
Consequently, starting from the half-filled system ({\it i.e.,} the uncompensated doped semiconductor), the system with a small percentage of extra electrons, because of the more extended wavefunction, experiences a greatly reduced effect of the positional disorder as compared with the corresponding hole-doped (less than half filled, {\it i.e.}~compensated) system, and so will behave more like the uniform lattice.  In diluted magnetic semiconductors (DMS), the existence of the relatively larger Bohr radius of the carriers ($\approx$10\AA) compared to the extent of the spins on the magnetic ions (1-2\AA) allows the carrier-magnetic moment interaction to dominate, resulting in a ferromagnetic ground state\cite{Berciu_DMS}.  In the electron-doped semiconductor, electrons occupying the $D^-$ state have a larger Bohr radius than the electrons giving rise to the effective exchange interaction $(J\sim\tInner^2/U)$.  This could cause carrier hopping to dominate, similarly resulting in a ferromagnetic ground state.  
At the very least, the different radii of the doubly vs. singly occupied sites suggests that we modify the Hamiltonian (\ref{eqnHubHamDisordered}) to become:
\begin{equation}
\mathcal{H} = - \sum_{\langle i,j\rangle\sigma} \left( t_{ij}(n_i,n_j)c^\dag_{i\sigma} c_{j\sigma} + \mbox{c.c} \right) + U\sum_i n_{i\uparrow}n_{i\downarrow} \label{eqnHubHamDisorderedOccDep}
\end{equation}
where $n_i$ is the total occupation of site $i$, and $t_{ij}$ is now has an occupation dependence given by:
\begin{displaymath}
t_{ij}(n_i,n_j) = \left\{ \begin{array}{cc}
\tOuter & n_j=2 \,\, \mbox{and}\,\, n_i=1 \label{tOuterCond} \\
\tInner & \mbox{otherwise} 
\end{array} \right.
\end{displaymath}
where $\tOuter$ is larger (and can be much larger) than $\tInner$.  For greater than one electron per site, the low energy spectrum in the limit $U\gg \tOuter,\tInner$ is given by the $\tOuter-J$ model, where $J=4\tInner^2/U$, as one would expect\cite{ChernyshevEffTheories}.  Hirsch has investigated a similar Hubbard model with occupation-dependent hopping, but  with focus on its prediction of superconductivity pairing\cite{HirschOccDepHopping}.  We proceed with semiconductors in mind, motivated by the notion that one might expect to find a ferromagnetic phase in this model.

\begin{figure}[H]
\begin{center}
\includegraphics[width=2.5in]{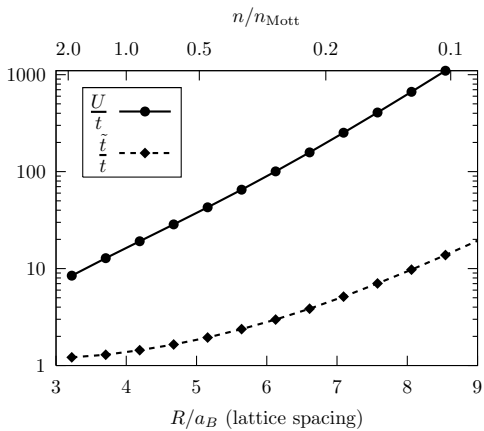}
\caption{Variation of ratios $U/t$ and $\tOuter/\tInner$ with the dopant spacing (related to the dopant density $\rho$ by $\rho = \frac{1}{R^3}$.
\label{figParamRatiosVsLatSpacing}}
\end{center}
\end{figure}

In the present work, we focus on the effect of changing the ratio $\tOuter/\tInner$ on small random clusters and lattice systems.  We have computed $U/t$ and $\tOuter/\tInner$ appropriate for dopants in semiconductors by performing a realistic calculation of single particle states of donors placed on a simple cubic lattice.  We extract the dependence of these ratios on the simple cubic lattice constant  by fitting the single particle bands to a tight binding model. Details of this work will be given elsewhere\cite{ENielsen}; here we show the resulting parameter ratios in figure \ref{figParamRatiosVsLatSpacing}.  The Mott metal-insulator transition criterion in these units is $R/a_{\mathrm{B}}=4$.  We see clearly that the range of $U/t$ and $\tOuter/\tInner$ can be varied substantially in doped semiconductors.  The large span of $U/t$ originates in the exponential dependence of the hopping parameter on the atomic spacing, and the variation of $\tOuter/\tInner$ from the relatively large size of the two-electron wavefunction appearing as a factor in this exponential.  Here we restrict ourselves to ratios in the ranges: $U/t=[10,100]$ and $\tOuter/\tInner=[1,10]$, which are conservative when compared to the physically attainable ranges.  

\begin{figure}[H]
\begin{center}
\includegraphics[width=2.9in]{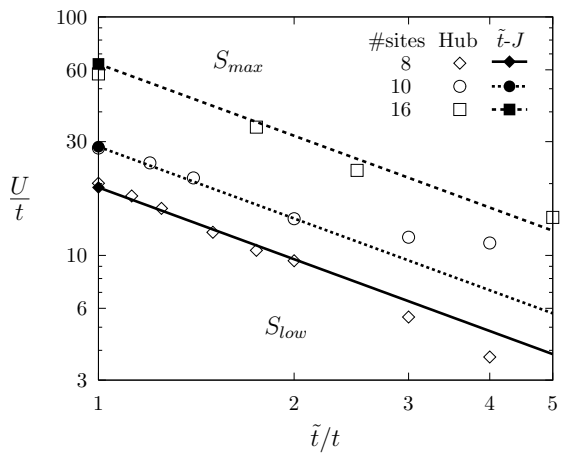}
\caption{Ground state spin diagram resulting from the exact diagonalization of (\ref{eqnHubHamDisorderedOccDep}) on 8-,10-, and 16-site square lattices with 9,11,and 17 electrons respectively.  Hubbard model results are displayed as open symbols.  Closed symbols and lines show the result of the corresponding $\tOuter-J$ model as described in the text. $S_{max}$ is the largest allowed spin on each lattice, and $S_{low}$ denotes a low ground state spin ($1/2$ for 10- and 16-site clusters, the minimum value).\label{figLatticeResults}}
\end{center}
\end{figure}

We have solved the Hubbard and corresponding $\tOuter$-$J$ models exactly using the Lanczos method to diagonalize the Hamiltonian matrix, after first taking advantage of all available spatial symmetries of the system and the spin symmetries (where allowed by memory constraints).  The Hubbard model depends on $U/\tInner$ and $\tOuter/\tInner$, whereas the $\tOuter-J$ model only depends on $\tOuter/J = \frac{1}{4}(\tOuter/\tInner)(U/\tInner)$.  Thus, the value of $\tOuter/J$ marking the onset of the Nagaoka state defines a line in $U/\tInner$ vs. $\tOuter/\tInner$ space.  Figure \ref{figLatticeResults} shows the ground state spin phase diagram for the 8-,10-,and 16-site square lattices doped with one extra electron.  One sees that the increase in $\tOuter/\tInner$ causes the area of the  maximal spin ground state phase to increase for electron-doped systems. 
We have also solved the Hubbard model on square lattices with a single hole, and see virtually no dependence of the ground state spin on $\tOuter/\tInner$, indicating a pronounced electron-hole asymmetry.  It is clear that this asymmetry originates from the electronic states having greater radius than the hole states, since the square lattice is bipartite and thus for $\tOuter=\tInner$ the problem is electron-hole symmetric.  In the case of a non-bipartite lattice or cluster, there is an asymmetry even in the simple Hubbard model\cite{MerinoTriangleLatPastorClusters}, and taking  $\tOuter>\tInner$ may yield a combined effect whereby the initial asymmetry is enhanced.

\begin{figure}[H]
\begin{center}
\includegraphics[width=3.2in]{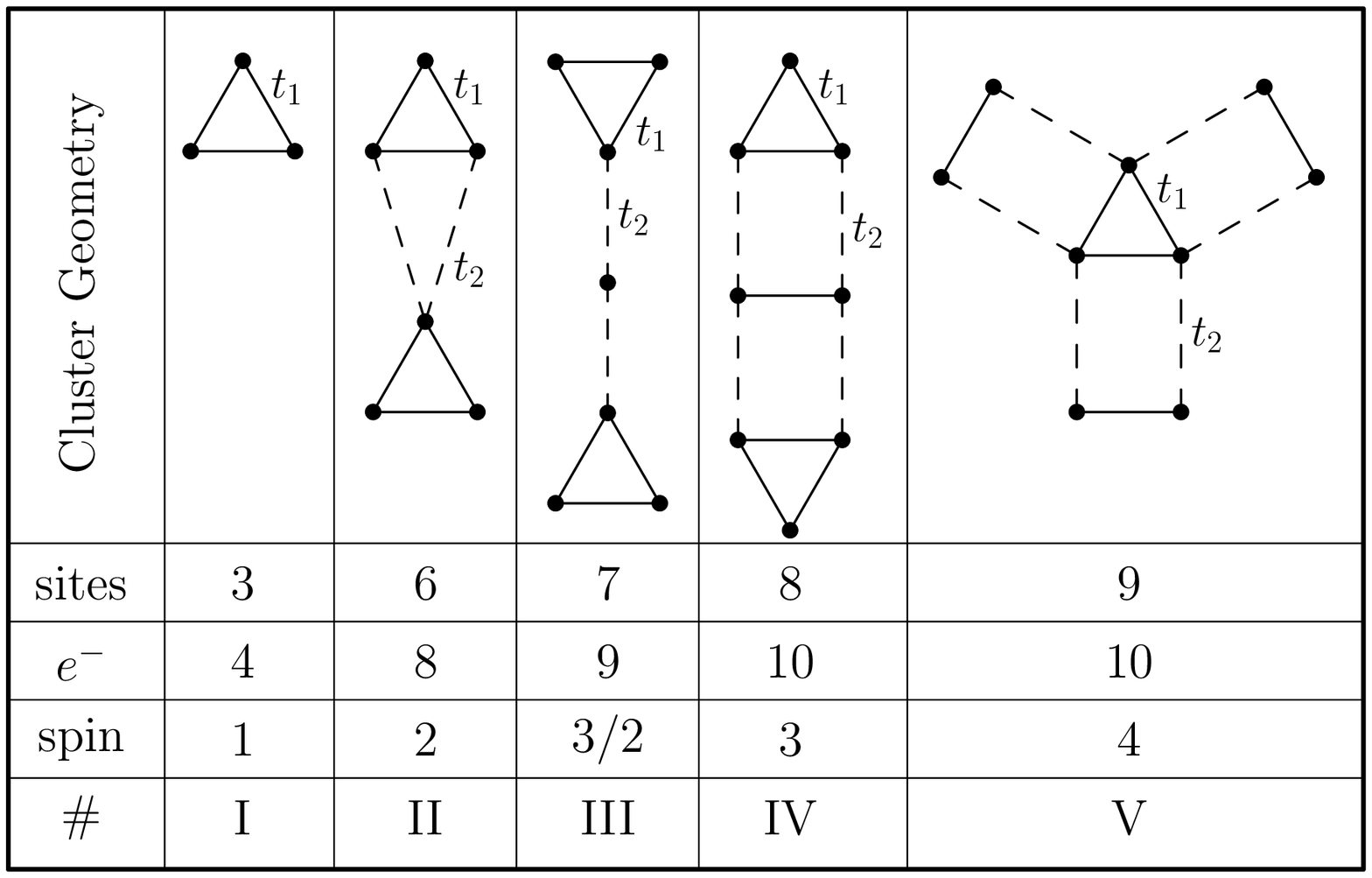}
\caption{Geometries of several high-spin clusters. The spin given is the ground state spin when $U/t_1 = 100$ and $t_2/t_1 = 0.5$.  These high spin ground states persist in considerable regions of phase space around this point.
\label{figClusterGeometries}}
\end{center}
\end{figure}

\begin{figure}
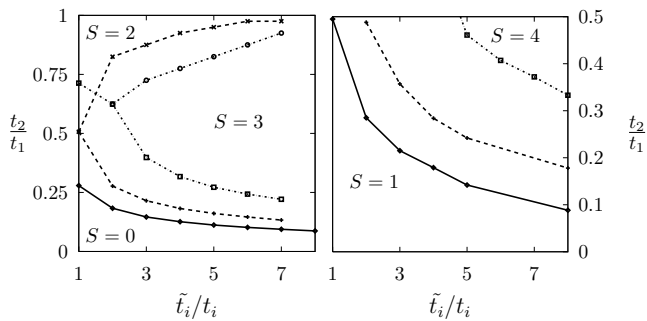
 
\begin{center}
\includegraphics[width=1.63in]{pairLinkedTriangles2.ps} 
\includegraphics[width=1.68in]{triangleInHex1.ps}
\caption{Ground state spin diagram for cluster IV (left) and V (right) of fig.~\ref{figClusterGeometries} with $10e^-$ as a function of $\tOuter/\tInner$.  Lines denote $U=100$ (solid), $U=50$ (dashed), and $U=20$ (dash-dot).\label{figClusterPhaseDiagram}}
\end{center}
\end{figure}

Figure \ref{figClusterGeometries} shows several clusters, each less symmetric than a lattice but retaining some spatial symmetries, that have a high-spin ground state in certain parameter ranges.  The magnetic phase diagram of the two largest clusters are shown in figure \ref{figClusterPhaseDiagram}, where $\tOuter/\tInner$ is plotted vs. a parameter measuring the geometry of the cluster.  We again see that the increase in $\tOuter/\tInner$ enlarges the high-spin region of the phase diagram and thus indicates that the high-spin state becomes more robust.

To further test the robustness of a cluster's high spin ground state, we multiply each $t_i$ by a random factor $\lambda$ whose logarithm is chosen from the box distribution $P(\log\lambda)=1/(2\log\alpha)$, $\log\lambda \in [-\log\alpha,+\log\alpha)$.  This more accurately characterizes the fluctuations we expect in a random system, since the hopping is exponentially dependent on the inter-site distance and we do not expect the fluctuations to preserve any symmetry present in the cluster.  We start with a cluster known to have a large spin ground state (\emph{e.g}~any cluster in fig.~\ref{figClusterGeometries}) and average over many of the just described random perturbations.  We find that the percentage of randomly perturbed clusters that \emph{retain} the high spin ground state of the original cluster increases dramatically with $\tOuter/\tInner$ (set to the same value on all edges).  Table \ref{tabPcRandomized} shows how this percentage depends on $\tOuter/\tInner$ and $\alpha$ for the particular cluster IV of fig.~\ref{figClusterGeometries}  with $t_1=1$, $t_2=0.3$, $U=100$. We see clearly that increasing $\tOuter/\tInner$ makes the high spin ground states of the clusters in fig.~\ref{figClusterGeometries} significantly more robust to geometric fluctuations present in the actual system.

\medskip

\begin{table}[H]
\begin{center}
\begin{tabular}{|c|c|c|c|}
  \hline & & & \\ [-2.2ex]
 \parbox[b]{2cm}{\centering$\tOuter/\tInner=$} & \parbox{1.5cm}{\centering 1.0} & \parbox{1.5cm}{\centering 2.5} & \parbox{1.5cm}{\centering 5.0} \\ [0.5ex] \hline
 $\alpha=0.7$ & 67 & 100 & 100 \\ 
 $\alpha=0.5$ & 26 & 86 & 90 \\ 
 $\alpha=0.3$ & 3 & 22 & 48 \\ \hline
\end{tabular}
\caption{Percentage of random instances of cluster IV (fig.~\ref{figClusterGeometries}) that retain a high spin ($S=3$) ground state.  Clusters are obtained by starting with $t_1=1$, $t_2=0.3$, $U=100$, and multiplying $\tInner$ and $\tOuter$ on each edge by a random factor in $[\alpha,1/\alpha]$. \label{tabPcRandomized}}
\end{center}
\end{table}

In conclusion, we have shown that on both finite (periodic) lattices and several less symmetric clusters, increasing $\tOuter/\tInner$ makes the appearance of nanoscale ferromagnetism significantly more likely.  The effect is greater for large $U/t$, which implies small dopant densities, well below the metal-insulator transition density. 
Such high spin states should be observable in doped quantum dots with dopant number $N_d = 6-15$ and with a small excess of electrons $N_e-N_d=1-2$.  The same density and (excess) electron doping regime is also the most likely region for the possible appearance of true macroscopic ferromagnetism,{\it e.g.}~in modulated structures with dopants in both quantum wells and barrier regions,
so that the region of excess electrons can be achieved, unlike in a true bulk doped semiconductor. 
However, obtaining a conclusive answer to this question numerically requires going beyond the small sizes possible with exact diagonalization methods, using, {\it e.g.}, density matrix or perturbative renormalization group methods, in combination with other numerical techniques.
Even if true ferromagnetism on the macroscopic scale is absent, our calculations show that there should be a significant asymmetry between the magnetic response of systems with excess electrons above the half-filled (uncompensated) case, and those with a deficit of electrons from the half-filled case, ({\it i.e.}~traditional compensated): the former should have a larger susceptibility in the paramagnetic phase at low temperatures. 
This can be experimentally checked by using gates to tune the electron density.
If ferromagnetism is attained on large enough length scales, it may show up as hysteresis in transport measurements due to magnetic domains.

This research was supported by NSF-MRSEC, Grant DMR-0213706.


\end{document}